\documentclass[11pt]{article}

\parskip \medskipamount

\usepackage[a4paper, left=2.5cm, right=2.5cm, top=1.8cm, bottom=1.8cm, includehead, includefoot, head=30pt]{geometry} 	% Manueel de grootte van de bladvulling bepalen
\usepackage{amsfonts}
\usepackage{natbib}											% Voor mooie citaties met \citet{}
\usepackage{graphicx} 									% Om figuren te kunnen verwerken (tip: minder plaats met wrapfigure)
\usepackage[small,bf,hang]{caption}			% Voor iets mooiere captions bij figuren
\usepackage{subcaption}									% Voor meerdere kleine figuren op te nemen in 1 figuur via \begin{subfigure}
\usepackage{wrapfig}										% Figuren tussen tekst m.b.v. wrapfigure
\usepackage{multirow}
\usepackage[table]{xcolor}										% Om in een tabel te spreiden over meerdere rijen
\usepackage[latin1]{inputenc}						% Om niet ascii karakters rechtstreeks te kunnen typen
\usepackage{amsmath}										% Uitgebreide wiskundige mogelijkheden
\allowdisplaybreaks[1]									% allow page breaks in equations
\usepackage{amssymb}										% Voor R, C, N via mathbb{R}
\usepackage{amsthm}											% Voor theorems en proofs
\usepackage{booktabs}										% Voor tabellen
\usepackage{url}                        % Om url's te verwerken
\usepackage{eurosym}                    % om het euro symbool te krijgen
\usepackage{textcomp}                   % Voor onder andere graden celsius
\usepackage{fancyhdr}                   % Voor fancy headers en footers
\usepackage{listings} 									% Code in file
\usepackage{pdfpages}										%\includepdf
\usepackage{floatrow}
\usepackage[title,titletoc,toc]{appendix}				% \begin{appendices}
\usepackage{enumerate}									% Voegt een optionele parameter voor layout toe aan enumerate,
\usepackage[multiple]{footmisc}																		% bvb: 																			
\usepackage{enumitem}										% minder witruimte tussen items bij itemize
\usepackage{mathtools}										% voor := en =:
\usepackage{IEEEtrantools}							% IEEEeqnarray environment
\usepackage{authblk}											% use author blocks
\usepackage{tikz}												% Tekenen van figuren in latex
\usetikzlibrary{arrows,decorations.pathmorphing,backgrounds,positioning,fit,petri}
\usepackage[english]{babel}
\usepackage{grffile}
\usepackage{amsfonts}
\usepackage{amssymb}
\usepackage{rotating}
\usepackage{siunitx}
\usepackage{color}	% Wiskunde in kleur via $\color{red} \alpha$
\usepackage{etoolbox}
\usepackage{todonotes}
\usepackage[plainpages=false,hyperfootnotes=false]{hyperref}
\hypersetup{
	colorlinks   = true, %Colours links instead of ugly boxes
	urlcolor     = blue, %Colour for external hyperlinks
	linkcolor    = red, %Colour of internal links
	citecolor   = blue %Colour of citations
}
\usepackage{kbordermatrix}% http://www.hss.caltech.edu/~kcb/TeX/kbordermatrix.sty
\usepackage[capitalize]{cleveref}

\usepackage{bookmark}
\usepackage[ruled,vlined,linesnumbered]{algorithm2e}
\SetAlCapFnt{\small} % Change the caption label font for the algorithms
\SetKw{KwTo}{in}

\SetKwInput{kwModel}{Model}
\SetKwInput{kwInit}{Initialization}

\makeatletter

\pretocmd{\NAT@citex}{%
	\let\NAT@hyper@\NAT@hyper@citex
	\def\NAT@postnote{#2}%
	\setcounter{NAT@total@cites}{0}%
	\setcounter{NAT@count@cites}{0}%
	\forcsvlist{\stepcounter{NAT@total@cites}\@gobble}{#3}}{}{}
\newcounter{NAT@total@cites}
\newcounter{NAT@count@cites}
\def\NAT@postnote{}

\def\NAT@hyper@citex#1{%
	\stepcounter{NAT@count@cites}%
	\hyper@natlinkstart{\@citeb\@extra@b@citeb}#1%
	\ifnumequal{\value{NAT@count@cites}}{\value{NAT@total@cites}}
	{\ifNAT@swa\else\if*\NAT@postnote*\else%
		\NAT@cmt\NAT@postnote\global\def\NAT@postnote{}\fi\fi}{}%
	\ifNAT@swa\else\if\relax\NAT@date\relax
	\else\NAT@@close\global\let\NAT@nm\@empty\fi\fi% avoid compact citations
	\hyper@natlinkend}
\renewcommand\hyper@natlinkbreak[2]{#1}

\patchcmd{\NAT@citex}
{\ifNAT@swa\else\if*#2*\else\NAT@cmt#2\fi
	\if\relax\NAT@date\relax\else\NAT@@close\fi\fi}{}{}{}
\patchcmd{\NAT@citex}
{\if\relax\NAT@date\relax\NAT@def@citea\else\NAT@def@citea@close\fi}
{\if\relax\NAT@date\relax\NAT@def@citea\else\NAT@def@citea@space\fi}{}{}

\makeatother

\definecolor{darkgreen}{RGB}{40,150,40}

\newcommand{\bs}{\boldsymbol}

\renewcommand{\epsilon}{\varepsilon}

\theoremstyle{definition}

\setcitestyle{authoryear,open={(},close={)}}

\newcolumntype{K}{>{\centering\arraybackslash}m{1.5cm}}
\newcolumntype{L}{>{\centering\arraybackslash}m{2.5cm}}
\newcolumntype{M}{>{\centering\arraybackslash}m{3.5cm}}
\newcolumntype{A}{>{\raggedright\arraybackslash}m{2cm}}
\newcolumntype{B}{>{\raggedright\arraybackslash}m{5cm}}
\newcolumntype{N}{>{\raggedright\arraybackslash}p{9.0cm}}
\newcolumntype{C}[1]{>{\centering\let\newline\\\arraybackslash\hspace{0pt}}m{#1}}
\newcolumntype{D}{>{\raggedright\arraybackslash}m{1cm}}
\newcolumntype{H}{>{\setbox0=\hbox\bgroup}c<{\egroup}@{}}

\usepackage{threeparttable}
\newfloatcommand{capbtabbox}{table}[][\FBwidth]
\newfloatcommand{ImportTable}{input}[][\FBwidth]
\newfloatcommand{capthreepart}{threeparttable}[][\FBwidth]
\DeclareFloatFont{Small}{\fontsize{9}{11}\selectfont}
\DeclareFloatFont{Small2}{\small}
\DeclareFloatFont{footnote}{\footnotesize}
\DeclareFloatFont{scriptsize}{\scriptsize}
\DeclareMarginSet{hangboth}{\setfloatmargins*{\hskip-4cm}{\hskip-4cm}}

\DeclareNewFloatType{figurea}
{placement=htbp, name=Figure, within=section, fileext=lofa}

\newfloatcommand{ffigabox}{figurea}[\nocapbeside][]
\makeatletter
\def\thickhline{%
	\noalign{\ifnum0=`}\fi\hrule \@height \thickarrayrulewidth \futurelet
	\reserved@a\@xthickhline}
\def\@xthickhline{\ifx\reserved@a\thickhline
	\vskip\doublerulesep
	\vskip-\thickarrayrulewidth
	\fi
	\ifnum0=`{\fi}}
\makeatother

\newlength{\thickarrayrulewidth}
\newcolumntype{O}{>{\centering\arraybackslash}m{1.5cm}}

\let\origappendix\appendix % save the existing appendix command
\renewcommand\appendix{\clearpage\pagenumbering{roman}\origappendix}

\usepackage{subcaption}

\usepackage{nccmath}

\graphicspath{{Figures/}}

\author[1, 2]{Bavo De Cock Campo}
\affil[1]{Department of Development and Regeneration, Faculty of Medicine, KU Leuven, Belgium.}
\affil[2]{Insurance Research Group, Faculty of Economics and Business, KU Leuven, Belgium.}
\title{\textbf{Towards reliable predictive analytics: a generalized calibration framework}}

\begin{document}
\newcommand{\form}[1]{\scalebox{1.087}{\boldmath{#1}}}

			\sloppy
			\maketitle
			
			\begin{abstract}
				\noindent
				Calibration is a pivotal aspect in predictive modeling, as it ensures that the predictions closely correspond with what we observe empirically. The contemporary calibration framework, however, is predominantly focused on prediction models where the outcome is a binary variable. We extend the logistic calibration framework to the generalized calibration framework which includes all members of the exponential family of distributions. We propose two different methods to estimate the calibration curve in this setting, a generalized linear model and a non-parametric smoother. In addition, we define two measures that summarize the calibration performance. The generalized calibration slope which quantifies the amount of over- or underfitting and the generalized calibration slope or calibration-in-the-large that measures the agreement between the global empirical average and the average predicted value. We provide an illustrative example using a simulated data set and hereby show how we can utilize the generalized calibration framework to assess the calibration of different types of prediction models.
				
				\vspace{2mm}
			\end{abstract}

			%% 1. Introduction %%
			\section{Introduction}
% Importance calibration in prediction models --> no measure for distributions other than logistic --> extension.
Prediction models are omnipresent in today's scientific literature. To construct a prediction model, we may either rely on statistical or machine learning techniques \citep{Steyerberg2010, Wessler2017, Kleinrouweler2016}. Within biomedical research, prediction models are typically developed to estimate a patient's risk of a disease or a future health state \citep{EvaSystematicReview}. Using the predictions, we aim to support the medical decision-making process and to improve patient care and health outcomes \citep{VanCalster2019,Speiser2021}.

Validation plays a crucial role in the development of a prediction model. The model's predictions have to be generalizable to the target population, extending beyond the training data \citep{Wallace2011,ClinicalPredictionModels}. Hereto, we evaluate the predictive performance of the model on a validation data set. There are two different aspects of a model's predictive performance: discrimination and calibration. The discriminatory performance refers to the model's ability to differentiate between high- and low-risk patients \citep{Harrell1996,Alba2017}. This is commonly the main focus when evaluating the performance of a prediction model \citep{Collins2014}. However, restricting the assessment to discriminatory ability neglects other crucial aspects of the model's predictive performance \citep{Alba2017,EvaSystematicReview}. Assessing the calibration of a model is essential, as it plays a pivotal role in achieving accurate and reliable model performance. With calibration we refer to the agreement between the observed outcomes and predictions \citep{ClinicalPredictionModels,Alba2017,CalibrationML}. Even though good calibration is an ubiquitous aspect of prediction models \citep{Kyung2011, Vach2013, Pepe2013}, most studies do not assess this aspect \citep{EvaSystematicReview,Collins2014,CalibrationML}. Moreover, discrimination and calibration are two distinct facets of model performance. Good discrimination does not necessarily imply good calibration. It may well be that the model is able to discriminate between high- and low-risk patients even though the estimated risks are unreliable \citep{VanCalster2019}. This phenomenon is not limited to classical statistical models, but extends to machine learning models as well \citep{EvaSystematicReview,CalibrationML}. Poorly calibrated probabilities can lead to misleading interpretations and inaccurate decision-making. Consequently, it is imperative to evaluate the calibration of the predicted probabilities to assess the model's reliability and accuracy.

To assess the calibration of a risk prediction model for binary outcomes, several measures are available. One of the primary measures is the calibration plot or calibration curve, which visually depicts the agreement between predicted and observed probabilities \citep{Stevens2020,ClinicalPredictionModels,VanCalster2019,CalHierarchy}. When a model is perfectly calibrated, the predicted risks coincide with the observed proportions and in the plot, this results in a diagonal line. To estimate the calibration curve, three distinct methods can be employed. In the first method, we group the estimated risks in bins and calculate the average per bin. Hereafter, we plot the average predicted risk versus the observed event rate per bin. For the second method, we rely on logistic regression to estimate the calibration curve \citep{Cox1958,Miller1993}. In this framework, we also compute the calibration intercept and calibration slope which summarize the calibration performance. The calibration intercept or calibration-in-the-large indicates whether the risks are under- or overestimated on average. The calibration slope quantifies the degree of over- or underfitting. The third method relies on non-parametric smoothers to estimate the calibration curve and is commonly referred to as the flexible calibration curve \citep{CalHierarchy, VanCalster2019}.

Notwithstanding, most existing methods to assess the calibration are applicable only to risk prediction models with a Bernoulli response variable and its related forms, such as binomial and multinomial distributed variables. To the best of our knowledge, there is no existing framework for other types of variables. In this paper, we extend the logistic calibration framework to include a wider range of prediction models. More specifically, we define calibration for any type of outcome variable that follows a distribution of the exponential family. We propose procedures to estimate the calibration curve, intercept and slope. Using this framework, we are able to assess the calibration of prediction models for count data, for example.

The remainder of this paper is organized as follows. In \Cref{sec:methodology}, we describe the calibration framework as developed for risk prediction models with a binary outcome variable. We then present the generalized calibration framework, which encompasses all distributions that are part of the exponential family. We provide a practical demonstration of the generalized calibration framework in \Cref{sec:illustration}. We conclude with a discussion in \Cref{sec:discussion}.

			%% 2. Overview %%
			\section{Assessing the predictive accuracy using calibration curves}\label{sec:methodology}

\subsection{Risk prediction models}
Using risk prediction models, we estimate the probability $\pi_i$ of observing an event. We use $y_i \in (0, 1)$ to denote the variable that captures this outcome and which takes on the value 0 in case of a non-event and 1 in case of an event. Here, $i$ serves as an index for the observations with $i = (1, \dots, n)$ where $n$ denotes the total number of observations. We assume that the response variable $y_i$ follows a Bernoulli distribution $y_i \sim \text{Bern}(\pi_i)$.

We commonly assume that $\pi_i$ depends on a set of risk characteristics, contained in covariate vector\footnote{We assume all vectors to be column vectors, unless explicitly transposed.} $\boldsymbol{x}_i$, and that there exists an unknown regression function $r(\bs{x}_i) = pr(y_i = 1 | \boldsymbol{x}_i)$. We approximate this unknown function using a risk prediction model, where we model the outcome as a function of the observed risk characteristics. To develop a risk prediction model, we rely on statistical or machine learning techniques. A general expression that encompasses both types is
\begin{align}\label{eq:RiskPredModel}
	pr(y_i  = 1| \boldsymbol{x}_i) = \pi_i = f(\boldsymbol{x}_i).
\end{align}

\noindent
In a logistic regression model, this equation takes on the following form
\begin{align}
	\nonumber
	pr(y_i = 1| \boldsymbol{x}_i) = \frac{e^{\boldsymbol{x}_i^\top \boldsymbol{\beta}}}{1 + e^{\boldsymbol{x}_i^\top \boldsymbol{\beta}}}
\end{align}
where $\boldsymbol{\beta}$ denotes the parameter vector. We can rewrite the equation to its more well-known form
\begin{equation}\label{eq:LR}
	\begin{aligned}
		\log\left( \frac{pr(y_i  = 1| \boldsymbol{x}_i)}{1 - pr(y_i  = 1| \boldsymbol{x}_i)} \right) &= \boldsymbol{x}_i^\top \boldsymbol{\beta}\\[0.5em]
		\text{logit}(pr(y_i  = 1| \boldsymbol{x}_i)) &= \eta_i
	\end{aligned}
\end{equation}
where $\eta_i$ denotes the linear predictor.

With machine learning methods, the functional form $f(\cdot)$ in equation \eqref{eq:RiskPredModel} depends on the specific algorithm. Using tree-based methods, for example, this corresponds to the observed proportion in the leaf nodes. For neural networks, $f(\cdot)$ is determined by the weights in the layers and the chosen activation functions.

We fit the statistical or machine learning model on a training set to estimate equation \eqref{eq:RiskPredModel}. Using the fitted model we obtain the predicted probability $\widehat{\pi}_i = \widehat{f}(\bs{x}_i)$. We assume that we also have a validation set, on which we validate the model and its predictions. To differentiate between the training and validation set, we append the subscript $*$ to the quantities of the validation set. Hence, ${}_{*} y_i$ denotes the $i^{th}$ outcome in the validation set. Similarly, we use ${}_{*} \boldsymbol{x}_i$ to denote the covariate vector for observation $i$ in the validation set. 
%We then calculate the linear predictor on the test set as
%\begin{align*}
%	{}_{*} \widehat{\eta}_i = {}_{*} \boldsymbol{x}_i^\top \widehat{\boldsymbol{\beta}}.
%\end{align*}

%\textcolor{red}{regression function p is called a calibration curve as it maps the predicted probabilities to the actual or relcalibrated even probabilities}
%
%\textcolor{red}{Harrell: we can estimate the true calibrated probabilities given the predicted probabilities by estimating the relationship between the predicted probabilities and $y_i$}

\paragraph*{Accuracy of the risk estimates} Calibration is an essential aspect of risk prediction models and ensures that the predicted probability corresponds to the actual probability of the event occurring. One way to examine the calibration of risk predictions, is by using calibration curves \citep{ClinicalPredictionModels, CalHierarchy, VanCalster2019, Dimitridias2022}. A calibration curve maps the predicted probabilities $f(\bs{x}_i)$ to the actual event probabilities $pr(y_i = 1| f(\bs{x}_i))$ and visualizes the correspondence between the model's predicted risks and the true probabilities. For perfectly calibrated predictions, the calibration curve equals the diagonal, i.e. $pr(y_i = 1 | f(\boldsymbol{x}_i)) = f(\boldsymbol{x}_i) \ \forall \ i$. 

In practice, we typically assess the model's calibration on a validation set. To estimate the calibration curve, we can employ three different methods \citep{CalHierarchy}. In the first method, we start by grouping the predicted risks in bins. Similar to the Hosmer-Lemeshow test \citep{Hosmer1997}, we can create the bins using deciles. We obtain the calibration curve by plotting the average predicted risk versus the observed event rate for each of the bins. For the second method, we rely on the logistic calibration framework \citep{Cox1958,Miller1993}. Here, we fit the following logistic regression model
\begin{align}\label{eq:CalibrationLogistic}
	\text{logit}(pr({}_{*} y_i = 1| {}_{*} \widehat{\pi}_i)) &= \alpha + \zeta \ \text{logit}({}_{*} \widehat{\pi}_i).
\end{align}
Note that $\text{logit}({}_{*} \widehat{\pi}_i) = {}_{*} \widehat{\eta}_i$ when ${}_{*} \widehat{\pi}_i$ is estimated using a logistic regression model (see \eqref{eq:LR}). Using \eqref{eq:CalibrationLogistic} we estimate the observed proportions as a function of the (logit transformed) predicted probabilities. The model is perfectly calibrated when $\alpha = 0$ and $\zeta = 1$. The value of the calibration slope $\zeta$ tells us whether the model is over- ($\zeta < 1$) or underfitted ($\zeta >1$). When $\zeta < 1$, the $\widehat{\pi}_i$'s are too extreme and need to be lower to ensure that the predicted risks coincide with the observed risks. Here, we overestimate the high risks and underestimate the low risks. Conversely, when $\zeta > 1$, the predicted risks are not extreme enough. In the logistic calibration framework, we commonly also assess the calibration-in-the-large which evaluates whether the observed event rate equals the average predicted risk. To calculate the calibration-in-the-large, we fix the calibration slope at $1$ and denote this as $\alpha|\zeta = 1$ or the short-hand notation $\alpha_c$. To estimate $\alpha_c$, we fit the model 
\begin{align}\label{eq:calintercept}
	\text{logit}(pr({}_{*} y_i = 1| {}_{*} \widehat{\pi}_i)) &= \alpha_c + \text{offset}(\text{logit}({}_{*} \widehat{\pi}_i))
\end{align}
where we enter $\text{logit}({}_{*} \widehat{\pi}_i)$ as an offset variable. Hereby, we fix $\zeta = 1$. The calibration intercept tells us whether the risks are overestimated $(\alpha_c < 0)$ or underestimated $(\alpha_c > 0)$ on average.

A third method to estimate the calibration curve relies on non-parametric smoothers such as locally estimated scatterplot smoothing (loess) or restricted cubic splines \citep{Copas1983Smooth, Austin2014, Harrell2015RMS, CalHierarchy, VanCalster2019}. We refer hereto as the flexible calibration curve.

\subsection{The generalized calibration framework}
The calibration framework, however, has only been defined for risk prediction models with a Bernoulli response variable and its related forms, such as binomial and multinomial distributed variables. To the best of our knowledge, this framework has not yet been extended to include other distributions as well. We propose an extension of the logistic calibration framework to distributions that belong to the exponential family with probability density function (pdf)
\begin{equation}
	\begin{aligned}
		\nonumber
		f(y_i; \theta_i, \phi, w_i) = \exp\left( \frac{y_i \theta_i - b(\theta_i)}{\phi} w_i  + c(y_i, \phi, w_i)\right).
	\end{aligned}
\end{equation}
\noindent
Here, $\theta_i$ is the natural parameter, $\phi$ the dispersion parameter and $w_i$ the weight. $b(\cdot)$ and $c(\cdot)$ are known functions. Similar to before, we assume that there is an unknown regression function $r(\boldsymbol{x}_i) = E(y_i | \bs{x}_i)$. To approximate this unknown function, we rely on prediction models with the following functional form
\begin{align}\label{eq:PredModel}
	E(y_i | \boldsymbol{x}_i) = \mu_i = f(\boldsymbol{x}_i).
\end{align}

\noindent
To estimate \eqref{eq:PredModel}, we can use a generalized linear model \citep{GLM}
\begin{equation}\label{eq:GLM}
	\begin{aligned}
		g(E(y_i | \boldsymbol{x}_i)) = \boldsymbol{x}_i^\top \boldsymbol{\beta} = \eta_i.
	\end{aligned}
\end{equation}
where $g(\cdot)$ denotes the link function. Alternatively, we can estimate \eqref{eq:PredModel} using machine learning methods. Using the model fit, we obtain the predictions $\widehat{\mu}_i = \widehat{f}(\boldsymbol{x}_i)$.

\paragraph*{Extending the calibration framework} To examine the calibration of prediction models where the outcome's distribution is a member of the exponential family, we redefine the framework in more general terms. In this context, a calibration curve maps the predicted values $f(\bs{x}_i)$ to $E(y_i| f(\bs{x}_i))$, the actual conditional mean of $y_i$ given $f(\bs{x}_i)$. As before, a model is perfectly calibrated if the calibration curve equals the diagonal, i.e. $E(y_i | f(\boldsymbol{x}_i)) = f(\boldsymbol{x}_i) \ \forall \ i$. In this context, the calibration curve captures the correspondence between the predicted values and the conditional mean. We refer to this curve as the generalized calibration curve. 

We propose two methods to estimate the generalized calibration curve. Firstly, we can estimate the calibration curve using a generalized linear model
\begin{equation}\label{eq:CalibrationGLM}
	\begin{aligned}
		g(E({}_{*} y_i | {}_{*} \widehat{\mu}_i)) = \alpha + \zeta \ g({}_{*} \widehat{\mu}_i).
	\end{aligned}
\end{equation}
By transforming ${}_{*} \widehat{\mu}_i$ using the appropriate $g(\cdot)$, we map ${}_{*} \widehat{\mu}_i$ to the whole real line to better fit the model. If ${}_{*} \widehat{\mu}_i$ is estimated using a generalized linear model with the same link function (i.e. $g(\cdot)$ is identical in \eqref{eq:GLM} and \eqref{eq:CalibrationGLM}), it follows that $g({}_{*} \widehat{\mu}_i) = {}_{*} \widehat{\eta}_i$. Using equation \eqref{eq:CalibrationGLM}, we estimate the empirical averages as a function of the predicted values. Further, similarly to \eqref{eq:CalibrationLogistic}, the generalized calibration slope $\zeta$ tells us whether the model is over- ($\zeta < 1$) or underfitted ($\zeta >1$). We estimate the generalized calibration intercept $\alpha_c$ as
\begin{equation}\label{eq:CITLGLM}
	\begin{aligned}
		g(E({}_{*} y_i | {}_{*} \widehat{\mu}_i)) = \alpha_c + \text{offset}(g({}_{*} \widehat{\mu}_i)).
	\end{aligned}
\end{equation}
Hereby, we assess to which extent the observed empirical average equals the average predicted value. Secondly, as with the risk prediction model, we can employ non-parametric smoothers to estimate the calibration curve.

			%% 3. Illustration %%
			\section{An empirical demonstration through synthetic data generation}\label{sec:illustration}
In this section, we illustrate how the generalized calibration framework can be employed to assess the calibration of prediction models for outcomes following an exponential family distribution. Hereto, we provide a case example where we assess the calibration of several prediction models developed for count data.

\paragraph*{Implementation} We perform the analysis using the statistical software \verb|R| \citep{Rsoftware}, version 4.2.3. The methods described in this article are implemented in the \href{https://cran.r-project.org/package=CalibrationCurves}{CalibrationCurves} package \citep{CalibrationCurves} which is available on the Comprehensive R Archive Network (CRAN) and on \href{https://github.com/BavoDC/CalibrationCurves}{Github}. The case example presented in this section, along with the corresponding analysis, can be replicated using the R script that is available on \url{https://github.com/BavoDC/PaperGeneralizedCalibrationCurves/EmpiricalDemonstration.R}.

\subsection{Assessing the calibration of a prediction model for count data}

\paragraph*{A synthetic data set}
To demonstrate the applicability of the generalized calibration framework, we generate a simulated data set. As such, we can provide a clear illustration of the method in a controlled environment where the underlying properties and relationships are known. To simulate the data, we first randomly generate five predictor variables $x_1$ to $x_5$. We take $N = 1\ 000 \ 000$ random draws from a multivariate standard normal distribution $\mathcal{N}(\bs{\mu}, \bs{\Sigma})$ where $\bs{\mu} = (0, 0, 0, 0, 0)$ and where the covariance matrix is defined as
\begingroup
\setlength\arraycolsep{4pt}
\begin{equation}
	\nonumber
	\Sigma = \begin{pmatrix}
		1.000 & 0.025 & 0.000 & 0.050 & 0.000 \\ 
		0.025 & 1.000 & 0.000 & 0.075 & 0.025 \\ 
		0.000 & 0.000 & 1.000 & 0.000 & 0.000 \\ 
		0.050 & 0.075 & 0.000 & 1.000 & 0.000 \\ 
		0.000 & 0.025 & 0.000 & 0.000 & 1.000 \\
	\end{pmatrix}.
\end{equation}
\endgroup

Hence, we introduce a small correlation between certain covariates to reflect a more realistic scenario. We rescale the standard normal variables such that $x_{ij} \in [-1, 1]$, where $i = (1, \dots, N)$ serves as an index for the observations and $j = (1, \dots, 5)$ as an index for the predictors. 

To simulate the outcome $y_i$, we employ the following data-generating model
\begin{equation}\label{eq:DataGeneratingModel}
	\begin{aligned}
		 E(y_i | \bs{x}_i) &= e^{\bs{x}_i^\top \bs{\beta}} = \lambda_i\\
	\end{aligned}
\end{equation}
\noindent
where $\bs{\beta} = (-2.3, 1.5, 2, -1, -2, -1.5)$ denotes the true parameter vector and $\bs{x}_i = (x_{i1}, x_{i2}, x_{i3}, x_{i4}, x_{i5})$ represents the covariate vector. We specify $y_i$ to follow a Poisson distribution and simulate $y_i$ by taking a random draw from Poi($\lambda_i$) for $i = (1, \dots, N)$. We refer to $\mathcal{D} = \{y_i, \bs{x}_i\}_{i = 1}^{N}$ as the population data set. To replicate the data-gathering process in real-life, we take a sample of size $5000$ from $\mathcal{D}$. We refer hereto as the training set ${}_{t} \mathcal{D}$ and use it to develop our model. Hereafter, we take a sample of size $1000$ from $\mathcal{D}$ which serves as the validation set ${}_{*} \mathcal{D}$. Using the model fit on ${}_{t} \mathcal{D}$, we predict the outcome in ${}_{*} \mathcal{D}$. We use the predictions in the validation set to assess the calibration of the model. We employ a Poisson generalized linear model with log link to estimate the generalized calibration curve, intercept and slope. Additionally, we estimate the generalized calibration curve using the non-parametric smoother loess.

%\textcolor{red}{Explain that we use a GAM and mention which smooth effect is fitted!}\\
To illustrate the generalized calibration framework, we fit three different models to ${}_{t} \mathcal{D}$
\begin{equation}\label{eq:FittedModels}
	\begin{aligned}
		&\hspace{-1.5cm}\text{generalized linear model 1}: \\
		&\hspace{2.5mm} \log({E(y_{i} | \bs{x}_i)}) = \bs{x}_i^\top \bs{\beta},\\
		&\hspace{-1.5cm}\text{generalized linear model 2}: \\
		&\hspace{2.5mm} \log({E(y_{i} | x_{i1}, x_{i3}, x_{i5})}) = x_{i1} \beta_1 + x_{i3} \beta_2 + x_{i1} x_{i3} \beta_3 + s(x_{i5}),\\
		&\hspace{-1.5cm}\text{generalized linear model 3}: \\
		&\hspace{2.5mm} \log({E(y_{i} | x_{i2})}) = x_{i2} \beta.
	\end{aligned}
\end{equation}
\noindent
where $s(\cdot)$ denotes a smooth effect. Model 1 represents the correctly specified model, whereas the other two models are incorrectly specified. Model 2 does not include variables $x_{2}$ and $x_{4}$. In addition, model 2 includes both an interaction term and a smooth effect which are not present in the data-generating model (see \eqref{eq:DataGeneratingModel}). Consequently, due to its complexity, this model is more prone to overfitting. Conversely, model 3 is oversimplified and includes only one variable. As such, it will not adequately capture the variation in the response and be more susceptible to underfitting. We rely on a Poisson generalized linear model with the log link function to fit models 1 and 3. To fit model 2, we use a Poisson generalized additive model \citep{gam, gamBook} with the log link function. Next, we use the fitted models to predict the outcome ${}_{*} y_i$ in the validation set ${}_{*} \mathcal{D}$. Additionally, since we know the true underlying data-generating model, we also calculate $E({}_{*} y_i | {}_{*} \bs{x}_i) = {}_{*} \lambda_i$.

\begin{figure}[!htbp]
	\centering
	\caption{\label{fig:PoissonCalibration}The generalized calibration curves, intercept and slope for: a) the true expected values ${}_{*} \lambda_i$; b) generalized linear model 1; c) generalized linear model 2 and d) generalized linear model 3. The dotted line represents the ideal line, the dashed line the calibration curve as estimated by a generalized linear model and the solid line depicts the flexible calibration curve. The grey area represents the pointwise 95\% confidence intervals of the flexible calibration curve. The histogram at the bottom of the plots shows the empirical distribution of the predicted values.}
	\makebox[\textwidth][c]{\includegraphics[scale = 0.5]{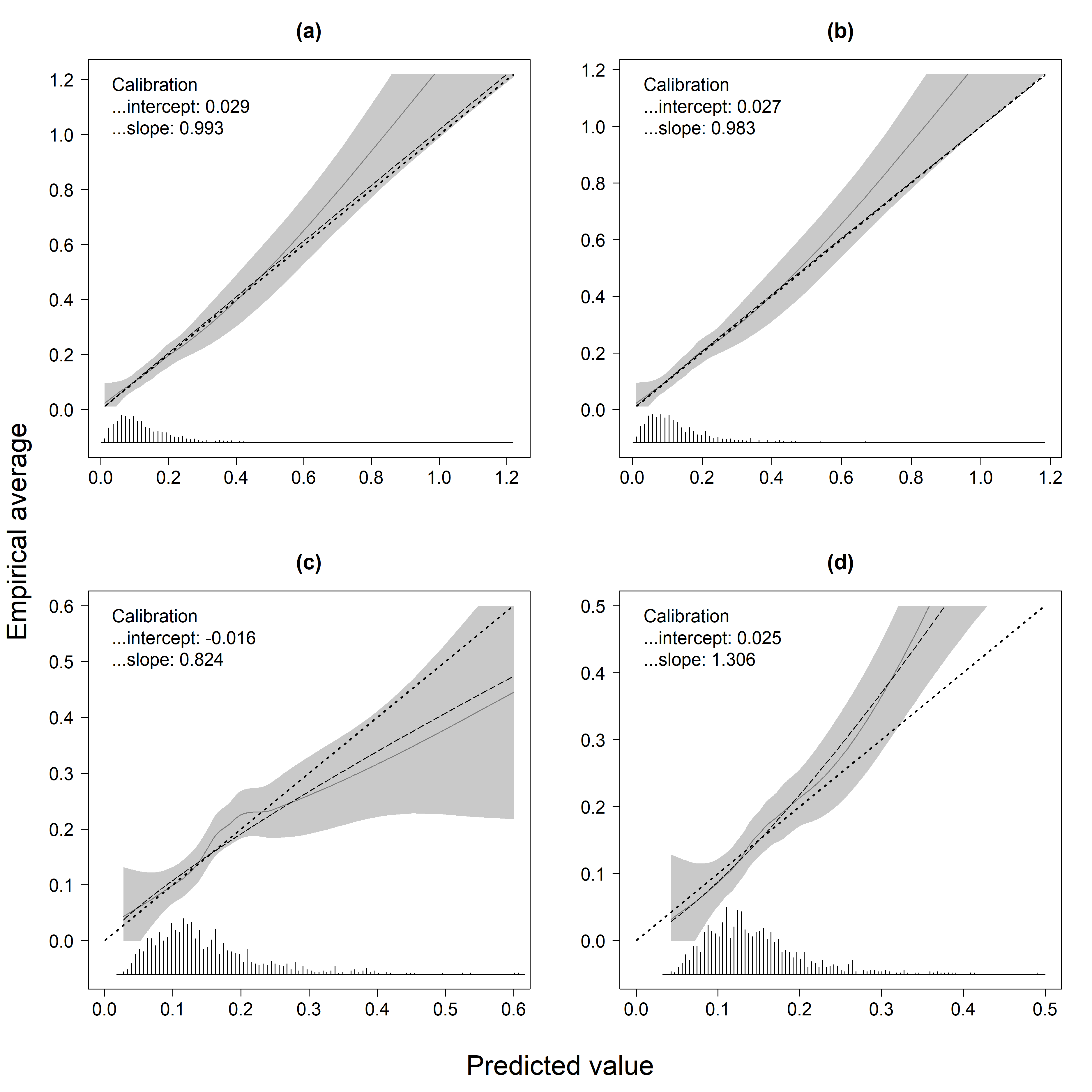}}
\end{figure}

\Cref{fig:PoissonCalibration} depicts the generalized calibration curve, intercept and slope for four different scenarios. Panel (a) depicts the calibration when we use the true ${}_{*} \lambda_i$'s. The remaining three panels show the calibration performance on ${}_{*} \mathcal{D}$  for the three fitted models. %In \Cref{fig:PoissonCalibration}, the diagonal red line represents the perfect calibration curve. The black dashed line and the black solid line represent the calibration curves as estimated by a generalized linear model and a smoother (loess), respectively. The grey area represents the pointwise 95\% confidence intervals of the flexible calibration curve. At the bottom of the plots, a histogram shows the empirical distribution of the predicted values. This provides an indication of the data volume or the sample size available. 

In real-world applications, we can only examine the correspondence between the predicted values and the empirical averages. As such, it provides an approximation to the model's calibration. It is important to acknowledge that the calibration evaluation is impacted by both the validation data set and the sampling variability. Consequently, attaining a model that demonstrates perfect calibration may prove to be unattainable in practice. We illustrate this by assessing the calibration using the ${}_{*} \lambda_i$'s (see \cref{fig:PoissonCalibration}(a)). The estimated calibration intercept and slope are only approximately equal to their ideal values ($\widehat{\alpha}_c \approxeq 0$ and $\widehat{\zeta} \approxeq 1$). In addition, the calibration curve as estimated by the generalized linear model does not perfectly coincide with the diagonal line. The flexible calibration curve is close to the diagonal line for low values of ${}_{*} \lambda_i$. For higher values, however, it deviates from the diagonal line. The histogram depicted at the bottom of the plot indicates that the data in this region is scarce. In addition, the pointwise confidence intervals are wide in this region. Hence, here we are unable to accurately assess the calibration due to insufficient data points. Moreover, it seems that the flexible calibration curve is drawn towards those few observations. Panel (b) depicts the calibration of generalized linear model 1 and shows a comparable pattern to panel (a). This model (see equation \eqref{eq:FittedModels}) is well-calibrated as the predicted values are approximately equal to the empirical averages. $\widehat{\alpha}_c \approxeq 0$, $\widehat{\zeta} \approxeq 1$ and the generalized linear model calibration curve coincides with the diagonal line. Similarly to panel (a), the flexible calibration curve deviates from the ideal line for higher values of the predicted values. We see a different pattern for generalized linear model 2, which is overfit to the data ($\widehat{\zeta} < 1$, see \cref{fig:PoissonCalibration}(c)). Low predicted values tend to (slightly) underestimate the empirical average whilst high predicted values overestimate the empirical average. \cref{fig:PoissonCalibration}(d) exhibits the opposite pattern. Generalized linear model 3 is underfit ($\widehat{\zeta} > 1$). We overestimate the empirical average for observations where the predicted value is low and underestimate the empirical average for observations where the predicted value is high.

\paragraph*{Expanding the scope: assessing the calibration of machine learning methods} Evaluating the calibration extends beyond traditional statistical methods and represents a pivotal aspect of any prediction model. We therefore illustrate how the generalized calibration framework can be applied to evaluate the calibration of predictions resulting from a machine learning model. As in the previous example, we estimate the generalized calibration curve, intercept and slope using a Poisson generalized linear model with log link. Additionally, we use loess as a non-parametric method to estimate the generalized calibration curve.

We opt for a gradient boosting machine \citep{Friedman2001} as our machine learning method of choice. This technique is a tree-based machine learning technique that combines multiple weak learners into a single powerful predictor. As loss function, we use the Poisson deviance. We set the minimum number of observations in the final node to be equal to 1\% of the total number of observations. For model training, we choose a shrinkage parameter value of 0.01, and during each iteration, we utilize 75\% of the training data as the proportion for updating the model. For gradient boosting machine 1 we perform an extensive grid search with the Poisson deviance as criterion to find the optimal values for the number of trees $T = (100, 200, \dots, 5000)$ and the tree depth $d = (1, 2, \dots, 10)$ using 10-fold cross-validation \citep{Hastie2009}. Based on this grid search, we select $T = 1600$ and $d = 1$ as the optimal parameters. All other parameter configurations are considered to be sub-optimal for this given data set. Based hereon, we employ two other parameter specifications. For gradient boosting machine 2, we let $T = 5000$ and $d = 5$. In this configuration we have an excessive number of trees and the depth of the trees goes beyond what is considered optimal. For gradient boosting machine 3,  we set $T = 200$ and $d = 1$. Hence, in this specification, we are working with an insufficient number of trees.

\begin{figure}[!htbp]
	\centering
	\caption{\label{fig:PoissonCalibrationGBM}The generalized calibration curves, intercept and slope of: a) gradient boosting machine 1 b) gradient boosting machine 2; and c) gradient boosting machine 3. The dotted line represents the ideal line, the dashed line the calibration curve as estimated by a generalized linear model and the solid line depicts the flexible calibration curve. The grey area represents the pointwise 95\% confidence intervals of the flexible calibration curve. The histogram at the bottom of the plots shows the empirical distribution of the predicted values.}
	\makebox[\textwidth][c]{\includegraphics[width = \textwidth]{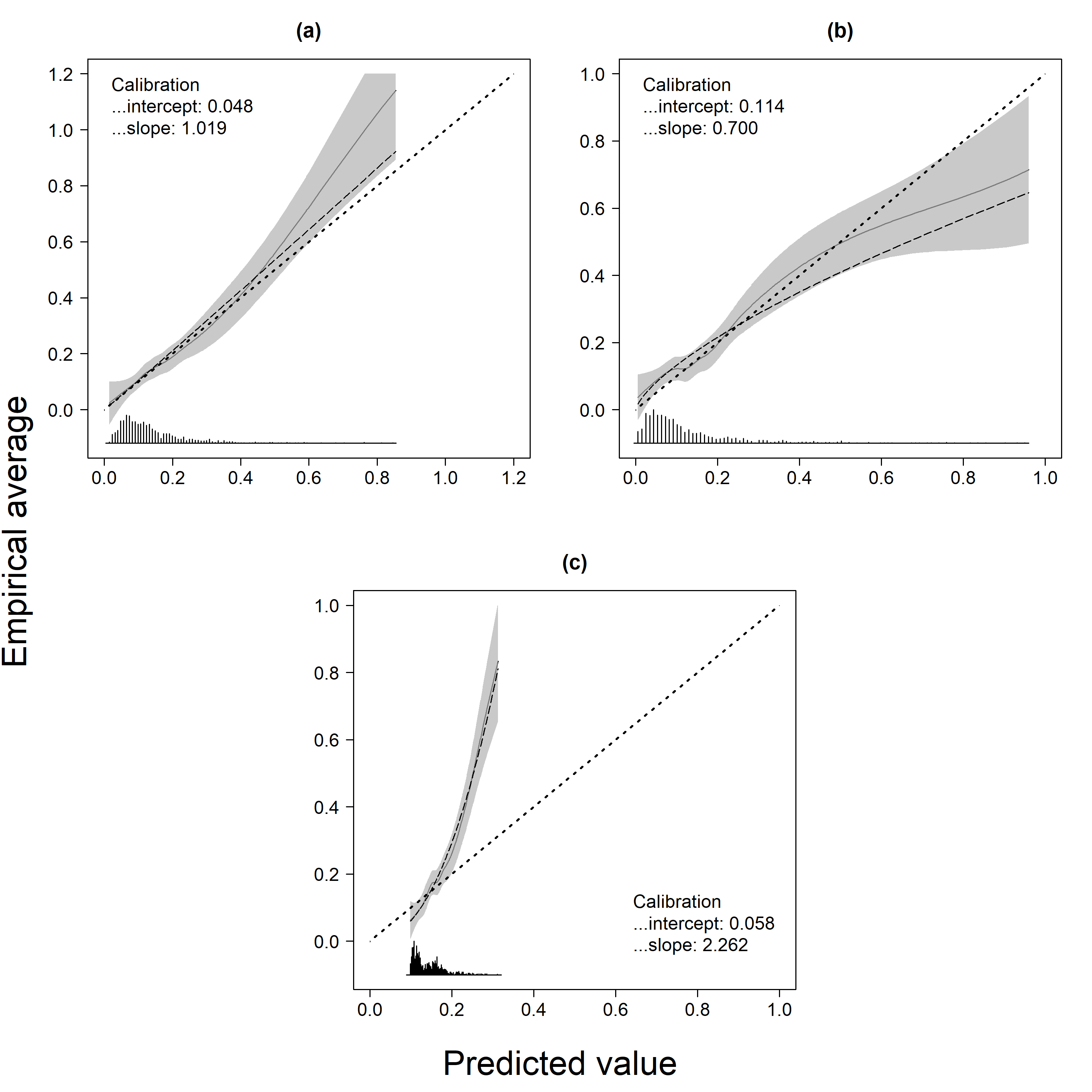}}
\end{figure}

We use the model fits to predict the outcome ${}_{*} y_i$ in the validation set ${}_{*} \mathcal{D}$ and \Cref{fig:PoissonCalibrationGBM} shows the calibration performance of the different gradient boosting machines. Gradient boosting machine 1 is the most well-calibrated model among all gradient boosting machines (see \cref{fig:PoissonCalibrationGBM}(a)). For this model, $\widehat{\alpha}_c \approxeq 0$ and $\widehat{\zeta} \approxeq 1$. Further, both the generalized linear model calibration curve and flexible calibration curve are close to the diagonal line. As with the correctly specified generalized linear model, the flexible calibration curve deviates from the ideal line for larger values of the predicted values where few observations are available. Panel (b) of \cref{fig:PoissonCalibrationGBM} depicts the calibration for gradient boosting machine 2. Similar to generalized linear model 2, gradient boosting machine 2 represents a typical example of a model that is overfit ($\widehat{\zeta} < 1$). This model underestimates the empirical average in cases where the predicted value is low and, conversely, overestimates the empirical average for observations characterized by high predicted values. In contrast, gradient boosting machine 3 illustrates a classic case of underfitting ($\widehat{\zeta} > 1$). The model tends to overestimate the empirical average in instances where the predicted value is low and this tendency shifts to underestimation for cases with higher predicted values. In addition, the predicted values have a limited range, with values spanning from 0.1 to 0.3. The model is too simple to capture the underlying patterns in the data and, consequently, is poorly calibrated.

			%% 3. Discussion and conclusion %%
			\section{Discussion}\label{sec:discussion}
In this paper, we set up a framework to assess the calibration of prediction models where the outcome follows a distribution from the exponential family. Hereto, we extend previously developed concepts of the logistic calibration framework to a more general setting. We propose two different methods to estimate the calibration curve, a generalized linear model and a non-parametric smoother. The estimated calibration curve gives us a graphical depiction of the correspondence between the predicted values and the empirical averages. In addition, we define two measures that summarize the calibration performance. The generalized calibration slope indicates whether the model is overfit or underfit. Models that are overfit provide predicted values that are too extreme. In cases with a low predicted value, we underestimate the empirical average. Conversely, for observations with a high predicted value, we overestimate the empirical average. We observe the opposite pattern for models that are underfit, where the predicted values are not extreme enough. The generalized calibration slope or calibration-in-the-large quantifies the agreement between the global empirical average and the average predicted value.

Using a simulated data set, we provide a practical example of how the generalized calibration framework can be employed to examine the calibration performance of a prediction model where the response variable is Poisson distributed. In our illustration, we develop several prediction models, using both statistical and machine learning techniques. We provide prototypical examples of models that are well-calibrated, overfit and underfit. %Additionally, we demonstrate that the precision of the calibration assessment is dependent on the amount of data we have available. %For certain predictions, we may have insufficient data to accurately assess the calibration. 

Future research can determine which method is best suited to estimate the calibration curve, what the drawbacks are of the different methods and how the estimation method interacts with the type of distribution of the response variable. The flexible calibration curve might be able to provide a better local approximation. However, non-parametric smoothers such as loess may exhibit erratic behavior in regions with sparse observations. Consequently, the flexible calibration curve is possibly less suited when the response variable's distribution is highly skewed.

			%% References %%
			\addcontentsline{toc}{chapter}{Bibliography}
			\bibliography{References}

\begin{thebibliography}{31}
\providecommand{\natexlab}[1]{#1}
\providecommand{\url}[1]{\texttt{#1}}
\expandafter\ifx\csname urlstyle\endcsname\relax
  \providecommand{\doi}[1]{doi: #1}\else
  \providecommand{\doi}{doi: \begingroup \urlstyle{rm}\Url}\fi

\bibitem[Alba et~al.(2017)Alba, Agoritsas, Walsh, Hanna, Iorio, Devereaux,
  McGinn, and Guyatt]{Alba2017}
A.~C. Alba, T.~Agoritsas, M.~Walsh, S.~Hanna, A.~Iorio, P.~J. Devereaux,
  T.~McGinn, and G.~Guyatt.
\newblock {Discrimination and Calibration of Clinical Prediction Models: Users'
  Guides to the Medical Literature}.
\newblock \emph{JAMA : the Journal of the American Medical Association},
  318\penalty0 (14):\penalty0 1377--1384, 2017.
\newblock ISSN 0098-7484.

\bibitem[Austin and Steyerberg(2014)]{Austin2014}
P.~C. Austin and E.~W. Steyerberg.
\newblock Graphical assessment of internal and external calibration of logistic
  regression models by using loess smoothers.
\newblock \emph{Statistics in Medicine}, 33\penalty0 (3):\penalty0 517--535,
  2014.
\newblock ISSN 0277-6715.

\bibitem[Christodoulou et~al.(2019)Christodoulou, Ma, Collins, Steyerberg,
  Verbakel, and Van~Calster]{EvaSystematicReview}
E.~Christodoulou, J.~Ma, G.~S. Collins, E.~W. Steyerberg, J.~Y. Verbakel, and
  B.~Van~Calster.
\newblock A systematic review shows no performance benefit of machine learning
  over logistic regression for clinical prediction models.
\newblock \emph{Journal of Clinical Epidemiology}, 110:\penalty0 12--22, 2019.
\newblock ISSN 0895-4356.

\bibitem[Collins et~al.(2014)Collins, de~Groot, Dutton, Omar, Shanyinde, Tajar,
  Voysey, Wharton, Yu, Moons, and Altman]{Collins2014}
G.~S. Collins, J.~A. de~Groot, S.~Dutton, O.~Omar, M.~Shanyinde, A.~Tajar,
  M.~Voysey, R.~Wharton, L.-M. Yu, K.~G. Moons, and D.~G. Altman.
\newblock External validation of multivariable prediction models: a systematic
  review of methodological conduct and reporting.
\newblock \emph{BMC Medical Research Methodology}, 14\penalty0 (1):\penalty0
  40--40, 2014.
\newblock ISSN 1471-2288.

\bibitem[Copas(1983)]{Copas1983Smooth}
J.~B. Copas.
\newblock Plotting p against x.
\newblock \emph{Journal of the Royal Statistical Society. Series C (Applied
  Statistics)}, 32\penalty0 (1):\penalty0 25--31, 1983.
\newblock ISSN 00359254, 14679876.
\newblock URL \url{http://www.jstor.org/stable/2348040}.

\bibitem[Cox(1958)]{Cox1958}
D.~Cox.
\newblock {Two further applications of a model for binary regression}.
\newblock \emph{Biometrika}, 45\penalty0 (3-4):\penalty0 562--565, 12 1958.
\newblock ISSN 0006-3444.
\newblock \doi{10.1093/biomet/45.3-4.562}.
\newblock URL \url{https://doi.org/10.1093/biomet/45.3-4.562}.

\bibitem[{De Cock} et~al.(2023){De Cock}, Nieboer, {Van Calster}, Steyerberg,
  and Vergouwe]{CalibrationCurves}
B.~{De Cock}, D.~Nieboer, B.~{Van Calster}, E.~W. Steyerberg, and Y.~Vergouwe.
\newblock \emph{The CalibrationCurves package: assessing the agreement between
  observed outcomes and predictions}, 2023.
\newblock URL \url{https://cran.r-project.org/package=CalibrationCurves}.
\newblock R package version 2.0.0.

\bibitem[Dimitriadis et~al.(2022)Dimitriadis, Dümbgen, Henzi, Puke, and
  Ziegel]{Dimitridias2022}
T.~Dimitriadis, L.~Dümbgen, A.~Henzi, M.~Puke, and J.~Ziegel.
\newblock {Honest calibration assessment for binary outcome predictions}.
\newblock \emph{Biometrika}, 110\penalty0 (3):\penalty0 663--680, 12 2022.
\newblock ISSN 1464-3510.
\newblock \doi{10.1093/biomet/asac068}.
\newblock URL \url{https://doi.org/10.1093/biomet/asac068}.

\bibitem[Dormann and Keil(2020)]{CalibrationML}
C.~F. Dormann and P.~Keil.
\newblock {Calibration of probability predictions from machine learning and
  statistical models}.
\newblock \emph{Global Ecology and Biogeography}, 29\penalty0 (4):\penalty0
  760--765, 2020.
\newblock ISSN 1466-822X.

\bibitem[Friedman(2001)]{Friedman2001}
J.~H. Friedman.
\newblock Greedy function approximation: A gradient boosting machine.
\newblock \emph{The Annals of statistics}, 29\penalty0 (5):\penalty0
  1189--1232, 2001.
\newblock ISSN 0090-5364.

\bibitem[Harrell et~al.(1996)Harrell, Lee, and Mark]{Harrell1996}
F.~E. Harrell, K.~L. Lee, and D.~B. Mark.
\newblock Multivariable prognostic models : Issues in developing models,
  evaluating assumptions and adequacy, and measuring and reducing errors.
\newblock \emph{Statistics in Medicine}, 15\penalty0 (4):\penalty0 361--387,
  1996.
\newblock ISSN 0277-6715.

\bibitem[Harrell(2015)]{Harrell2015RMS}
F.~E. Harrell, Jr.
\newblock \emph{Regression Modeling Strategies: With Applications to Linear
  Models, Logistic and Ordinal Regression, and Survival Analysis}.
\newblock Springer Series in Statistics. Springer International Publishing,
  Cham, 2nd ed. 2015. edition, 2015.
\newblock ISBN 3-319-19425-9.

\bibitem[Hastie et~al.(2009)Hastie, Tibshirani, and Friedman]{Hastie2009}
T.~Hastie, R.~Tibshirani, and J.~Friedman.
\newblock \emph{Elements of Statistical Learning: Data Mining, Inference, and
  Prediction}.
\newblock Springer series in statistics. Springer, New York, 2009.
\newblock ISBN 0387848576.

\bibitem[Hosmer et~al.(1997)Hosmer, Hosmer, Le~Cessie, and
  Lemeshow]{Hosmer1997}
D.~W. Hosmer, T.~Hosmer, S.~Le~Cessie, and S.~Lemeshow.
\newblock A comparison of goodness-of-fit tests for the logistic regression
  model.
\newblock \emph{Statistics in Medicine}, 16\penalty0 (9):\penalty0 965--980,
  1997.
\newblock ISSN 0277-6715.

\bibitem[Kim and Simon(2011)]{Kyung2011}
K.~I. Kim and R.~Simon.
\newblock Probabilistic classifiers with high-dimensional data.
\newblock \emph{Biostatistics (Oxford, England)}, 12\penalty0 (3):\penalty0
  399--412, 2011.
\newblock ISSN 1465-4644.

\bibitem[Kleinrouweler et~al.(2016)Kleinrouweler, Cheong-See, Collins, Kwee,
  Thangaratinam, Khan, Mol, Pajkrt, Moons, and Schuit]{Kleinrouweler2016}
C.~E. Kleinrouweler, F.~M. Cheong-See, G.~S. Collins, A.~Kwee,
  S.~Thangaratinam, K.~S. Khan, B.~W.~J. Mol, E.~Pajkrt, K.~G.~M. Moons, and
  E.~Schuit.
\newblock Prognostic models in obstetrics : available, but far from applicable.
\newblock \emph{American Journal of Obstetrics and Gynecology}, 214\penalty0
  (1):\penalty0 79--, 2016.
\newblock ISSN 0002-9378.

\bibitem[McCullagh and Nelder(1999)]{GLM}
P.~McCullagh and J.~A. Nelder.
\newblock \emph{Generalized linear models}.
\newblock Monographs on statistics and applied probability 37. Chapman and
  Hall, London, 2nd ed. repr. edition, 1999.
\newblock ISBN 0412317605.

\bibitem[Miller et~al.(1993)Miller, Langefeld, Tierney, Hui, and
  McDonald]{Miller1993}
M.~E. Miller, C.~D. Langefeld, W.~M. Tierney, S.~L. Hui, and C.~J. McDonald.
\newblock Validation of probabilistic predictions.
\newblock \emph{Medical Decision Making}, 13\penalty0 (1):\penalty0 49--57,
  1993.

\bibitem[Pepe and Janes(2013)]{Pepe2013}
M.~Pepe and H.~Janes.
\newblock Methods for evaluating prediction performance of biomarkers and
  tests.
\newblock In \emph{Risk Assessment and Evaluation of Predictions}, Lecture
  Notes in Statistics, pages 107--142. Springer New York, New York, NY, 2013.
\newblock ISBN 1461489806.

\bibitem[{R Core Team}(2023)]{Rsoftware}
{R Core Team}.
\newblock \emph{R: A Language and Environment for Statistical Computing}.
\newblock R Foundation for Statistical Computing, Vienna, Austria, 2023.
\newblock URL \url{https://www.R-project.org/}.

\bibitem[Speiser(2021)]{Speiser2021}
J.~L. Speiser.
\newblock So you developed a clinical prediction model, now what?
\newblock \emph{Journal of Data Science}, 19\penalty0 (4):\penalty0 519--527,
  2021.
\newblock ISSN 1680-743X.
\newblock \doi{10.6339/21-JDS1029}.

\bibitem[Stevens and Poppe(2020)]{Stevens2020}
R.~J. Stevens and K.~K. Poppe.
\newblock Validation of clinical prediction models: what does the "calibration
  slope" really measure?
\newblock \emph{Journal of Clinical Epidemiology}, 118:\penalty0 93--99, 2020.
\newblock ISSN 0895-4356.

\bibitem[Steyerberg(2019)]{ClinicalPredictionModels}
E.~W. Steyerberg.
\newblock \emph{Clinical Prediction Models: A Practical Approach to
  Development, Validation, and Updating}.
\newblock Statistics for biology and health. Springer International Publishing
  AG, Cham, 2019.
\newblock ISBN 3030163989.

\bibitem[Steyerberg et~al.(2010)Steyerberg, Vickers, Cook, Gerds, Gonen,
  Obuchowski, Pencina, and Kattan]{Steyerberg2010}
E.~W. Steyerberg, A.~J. Vickers, N.~R. Cook, T.~Gerds, M.~Gonen, N.~Obuchowski,
  M.~J. Pencina, and M.~W. Kattan.
\newblock Assessing the performance of prediction models: A framework for
  traditional and novel measures.
\newblock \emph{Epidemiology (Cambridge, Mass.)}, 21\penalty0 (1):\penalty0
  128--138, 2010.
\newblock ISSN 1044-3983.

\bibitem[Vach(2013)]{Vach2013}
W.~Vach.
\newblock Calibration of clinical prediction rules does not just assess bias.
\newblock \emph{Journal of Clinical Epidemiology}, 66\penalty0 (11):\penalty0
  1296--1301, 2013.
\newblock ISSN 0895-4356.

\bibitem[Van~Calster et~al.(2016)Van~Calster, Nieboer, Vergouwe, De~Cock,
  Pencina, and Steyerberg]{CalHierarchy}
B.~Van~Calster, D.~Nieboer, Y.~Vergouwe, B.~De~Cock, M.~J. Pencina, and E.~W.
  Steyerberg.
\newblock A calibration hierarchy for risk models was defined: from utopia to
  empirical data.
\newblock \emph{Journal of Clinical Epidemiology}, 74:\penalty0 167--176, 2016.
\newblock ISSN 0895-4356.

\bibitem[Van~Calster et~al.(2019)Van~Calster, McLernon, van Smeden, Wynants,
  Steyerberg, Bossuyt, Collins, MacAskill, Moons, and Vickers]{VanCalster2019}
B.~Van~Calster, D.~J. McLernon, M.~van Smeden, L.~Wynants, E.~W. Steyerberg,
  P.~Bossuyt, G.~S. Collins, P.~MacAskill, K.~G.~M. Moons, and A.~J. Vickers.
\newblock Calibration: The achilles heel of predictive analytics.
\newblock \emph{BMC Medicine}, 17\penalty0 (1):\penalty0 230--230, 2019.
\newblock ISSN 1741-7015.

\bibitem[Wallace et~al.(2011)Wallace, Smith, Perera-Salazar, Vaucher, McCowan,
  Collins, Verbakel, Lakhanpaul, and Fahey]{Wallace2011}
E.~Wallace, S.~M. Smith, R.~Perera-Salazar, P.~Vaucher, C.~McCowan, G.~Collins,
  J.~Verbakel, M.~Lakhanpaul, and T.~Fahey.
\newblock Framework for the impact analysis and implementation of clinical
  prediction rules (cprs).
\newblock \emph{BMC Medical Informatics and Decision Making}, 11\penalty0
  (1):\penalty0 62--62, 2011.
\newblock ISSN 1472-6947.

\bibitem[Wessler et~al.(2017)Wessler, Paulus, Lundquist, Ajlan, Natto, Janes,
  Jethmalani, Raman, Lutz, and Kent]{Wessler2017}
B.~S. Wessler, J.~Paulus, C.~M. Lundquist, M.~Ajlan, Z.~Natto, W.~A. Janes,
  N.~Jethmalani, G.~Raman, J.~S. Lutz, and D.~M. Kent.
\newblock Tufts pace clinical predictive model registry: update 1990 through
  2015.
\newblock \emph{Diagnostic and Prognostic Research}, 1\penalty0 (1):\penalty0
  20--20, 2017.
\newblock ISSN 2397-7523.

\bibitem[Wood(2017)]{gamBook}
S.~Wood.
\newblock \emph{Generalized Additive Models: An Introduction with R}.
\newblock Chapman and Hall/CRC, 2 edition, 2017.

\bibitem[Wood(2011)]{gam}
S.~N. Wood.
\newblock Fast stable restricted maximum likelihood and marginal likelihood
  estimation of semiparametric generalized linear models.
\newblock \emph{Journal of the Royal Statistical Society (B)}, 73\penalty0
  (1):\penalty0 3--36, 2011.

\end{thebibliography}
			%\bibliographystyle{ieeetr}	
			
			%%%%%%%%%%%%%%%%%%%%%%%%%%%%%%%%%%%%%%%%%%%%%%%%%%%%%%%%%%%%%%%%%%%%%%%%%%%%%%%%%%%%%%%%%%%%%%%%%%%%%%%%%%%%%%%%%%%%%%%%
			% Appendix
			\bookmarksetup{startatroot}
\end{document}